\documentclass[
reprint,
superscriptaddress,
amsmath,
amssymb,
]{revtex4-2}

\usepackage{graphicx}
\usepackage{dcolumn}
\usepackage{bm}
\usepackage{hhline}
\usepackage{array}
\usepackage{xcolor}
\usepackage{soul}
\usepackage[indent,skip=6pt]{parskip}

\begin{document}

\title{Microscopic Theory of Squeezed Light in Quantum Dot Systems}

\author{Sahil D. Patel}
\altaffiliation{These authors contributed equally to this work.}
\affiliation{Department of Electrical Engineering, University of California, Santa Barbara, USA}
\author{Sean Doan}
\altaffiliation{These authors contributed equally to this work.}
\affiliation{Department of Physics, University of California, Santa Barbara, USA}
\author{Chen Shang}
\affiliation{Department of Electrical Engineering, University of California, Santa Barbara, USA}
\author{Frederic Grillot}
\affiliation{Department of Electrical Engineering, University of California, Santa Barbara, USA}
\affiliation{Center for Optics, Photonics and Lasers, Laval University, Quebec City, Canada}
\affiliation{T\'el\'ecom Paris, Institut Polytechnique de Paris, Palaiseau, France}
\author{Frank Jahnke}
\affiliation{Institute for Theoretical Physics, University of Bremen, 28334 Bremen, Germany}
\author{John E. Bowers}
\author{Galan Moody}
\affiliation{Department of Electrical Engineering, University of California, Santa Barbara, USA}
\author{Weng W. Chow}
\affiliation{Sandia National Laboratories, Albuquerque, USA}
\date{\today}

\begin{abstract}
We present a cavity-QED theory for generating squeezed light from semiconductor quantum dots (QDs) integrated in microcavities. We formulate equations of motion for an inhomogeneously broadened QD ensemble that is incoherently pumped and simultaneously driven by a coherent seed field, solve for steady states, and compute the output-field quadrature variances. The analysis identifies operating conditions that yield amplitude-quadrature squeezing, with photon-number fluctuations reduced below the coherent-state limit and squeezing levels as large as 5 dB attainable with presently accessible QD and cavity parameters using only $\sim1$ $\mu$W pump power. We further show that quantum correlations originating from four-wave mixing play a dual role: they both shape the gain spectrum and generate squeezing. These correlations constitute the quantum counterpart of the mean-field (semiclassical) mechanisms responsible for self-mode-locking in QD lasers and the ultra-narrow lasing linewidths achieved under self-injection locking.
\end{abstract}

\maketitle

\section{Introduction}
\noindent Modern photonic systems rely on low-noise integrated lasers for precise control of amplitude and frequency content \cite{heindel2023quantum, bose2024anneal}, underpinning applications spanning optical timekeeping to LIDAR \cite{lukashchuk2024photonic, bianconi2025requirements}, data center interconnects \cite{fang2024overcoming, rizzo2023massively} and space-based communications \cite{zhang2021turbulence, zhu2021compensation}. For quantum information applications--where information is frequently encoded at the single-photon level--spontaneous emission imposes irreducible photon-number fluctuations that set the standard quantum limit (SQL) on the achievable signal-to-noise ratio \cite{andersen201630}. Operating below the shot-noise limit is therefore essential for tasks demanding enhanced precision \cite{aasi2013enhanced}] or quantum-verified security of optical links \cite{gehring2015implementation, gong2020secure}. 

In such scenarios, non-classical light is required to surpass the SQL. For these purposes, squeezed states are particularly useful, which are defined as when fluctuations in variance of one electromagnetic-field quadrature are reduced beneath the shot-noise level while noise in the conjugate quadrature is correspondingly increased, consistent with Heisenberg’s uncertainty principle \cite{walls1983squeezed}. Selecting whether amplitude or phase is compressed enables below-shot-noise measurements, yielding performance gains in quantum key distribution (QKD), continuous-variable quantum computing, and interferometric metrology \cite{aasi2013enhanced, zhong2021phase, lawrie2019quantum, derkach2020squeezing, nguyen2025digital}. Implementations in gravitational-wave detectors and secure fiber links show that even a several decibels of squeezing can deliver substantial improvements in sensing and communications \cite{tse2019quantum}. 

The primary methods for generating squeezed state of light rely on bulk nonlinear optics through spontaneous parametric down-conversion (in $\chi^{(2)}$ media) or spontaneous four-wave mixing (in $\chi^{(3)}$ media), using for example, lithium niobate (LiNbO$_3$) \cite{stefszky2017waveguide}, potassium titanyl phosphate (KTiOPO$_4$) \cite{vahlbruch2008observation}, silicon, silicon nitride (Si$_3$N$_4$) \cite{vaidya2020broadband}, and atomic vapors \cite{slusher1985observation, mccormick2006strong}. While up to $\sim15.7$ dB of squeezing has been demonstrated with bulk optics, these platforms typically depend on centimeter-scale resonators with hundreds of milliwatts of pump power and active thermo-mechanical stabilization. As a table-top apparatus, they can be challenging to scale, align, and deploy, thus limiting utilization in quantum systems with power and size constraints.

\begin{figure*}[t] \centering
\includegraphics[scale=0.7]{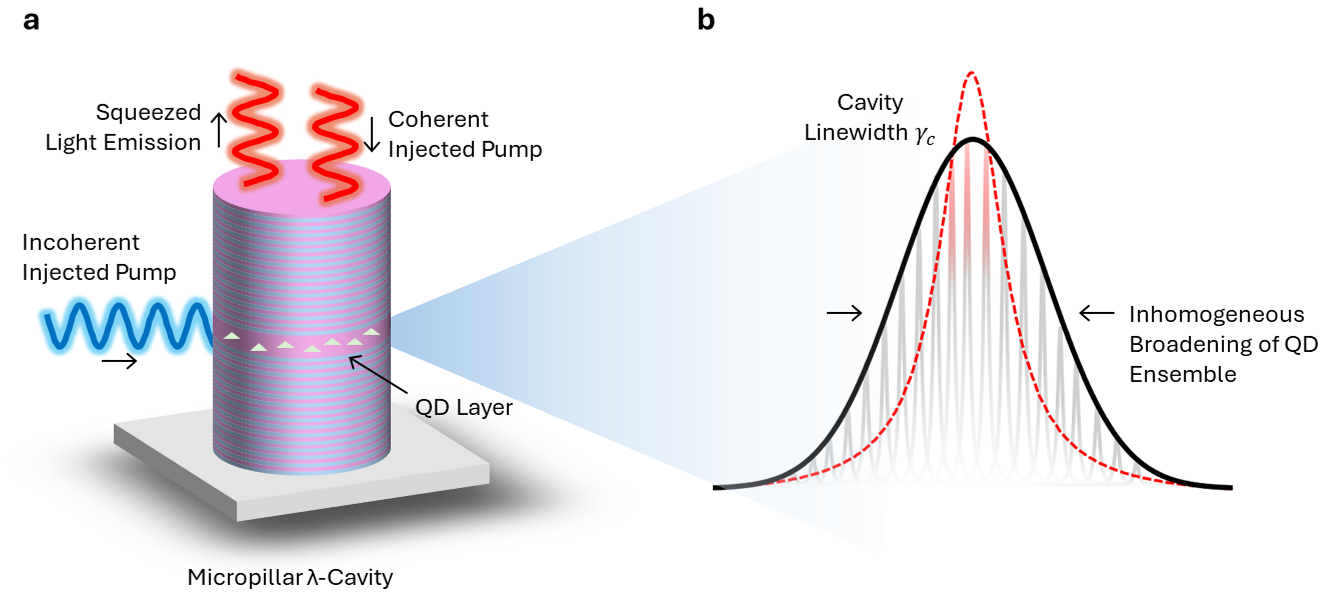}
\caption{\textbf{(a)} Schematic of a microcavity containing an inhomogeneously broadened QD ensemble that is incoherently pumped and driven by a coherent seed to produce squeezed light. \textbf{(b)} A subset of emitters couples to the cavity mode. For the parameters used here, 9/25 QDs contribute appreciably.}
\label{fig: sketch}
\end{figure*}

Semiconductor quantum dots (QDs) offer a compelling route to compact, integrable, and potentially room-temperature-compatible non-classical light sources \cite{heindel2023quantum, shang2025increasing}. Using lithographically defined photonic structures for coupling and routing, combined with growth and post-growth tuning (composition control, strain or Stark shift), QDs can be integrated with silicon photonics and positioned within telecom bands \cite{yu2023telecom}. Their discrete atom-like energy levels and large oscillator strengths, together with wafer-scale epitaxy, enable deterministic placement within microcavities where vacuum-field enhancement promotes strong light-matter coupling. When realized as lasers, high-density QD devices exhibit ultralow thresholds, robust thermal behavior, and ultrafast carrier dynamics, features associated with very low relative-intensity noise and sub-picosecond gain-recovery times \cite{grillot2024semiconductor, berg2002ultrafast}. Their pronounced third-order nonlinearity supports chip-scale four-wave mixing (FWM), and small cavity footprints enable close on-chip co-integration with control electronics and photonic waveguide networks.

Leveraging these advantages, room-temperature quadrature squeezing has recently been observed in QD lasers operated using low-noise electrical drive signals, reaching up to 3 dB of amplitude squeezing with $>$10-GHz measurement bandwidths \cite{zhao2024broadband}. The approach exploits cavity-enhanced FWM and carrier-population clamping to suppress amplitude noise without external nonlinear crystals or cryogenic cooling, highlighting the versatility of the QD platform. Together, these findings move the field beyond lab-bench optics toward monolithic squeezed-light sources compatible with standard semiconductor fabrication \cite{senellart2017high}. Such devices create opportunities for hybrid quantum-classical photonic circuits in which squeezed light, single-photon sources, and electronic drivers coexist on a single chip, enabling scalable quantum communications, sensing, and information processing.

In this work, we develop and apply a theory for squeezing in a microcavity that hosts an ensemble of inhomogeneously broadened QDs subject to incoherent pumping and a coherent injected seed field. Adopting a cavity quantum electrodynamics (cavity-QED) description, we derive equations of motion for the system and solve for steady state \cite{kreinberg2017emission}. We then evaluate the output-field quadrature variances. Operating regimes are identified that realize amplitude squeezing, with photon-number fluctuations reduced below the coherent-state (shot-noise) limit. We find that quantum correlations from FWM produce squeezing and reshape the gain profile, providing a quantum analogue of the mean-field processes responsible for self-mode-locking in QD lasers and the narrow linewidths seen under self-injection locking \cite{duan2022four, alkhazraji2023linewidth}. This framework extends semiclassical laser physics and introduces new design rules for integrated squeezed-light sources. By linking the microscopic dynamics of QD ensembles to the macroscopic cavity architecture, our model supports the engineering of compact, high-purity non-classical light sources and advances scalable quantum photonics and networking technologies.

\vspace{-0.3cm}

\section{Theory}

\begin{figure*}[t] \centering
     \includegraphics[scale=1.1]{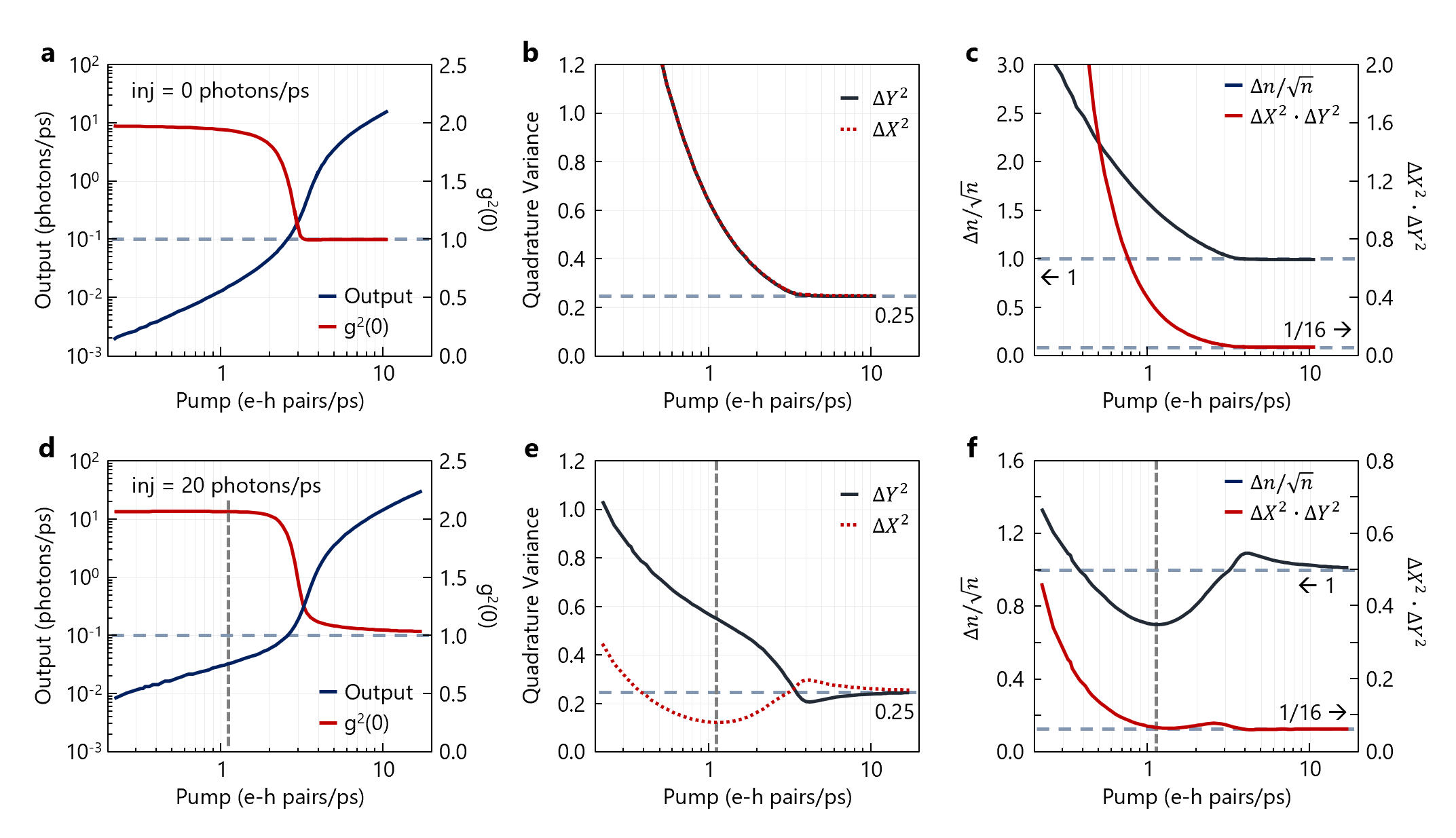}  
          \caption{\footnotesize Top (bottom) row, without (with) laser injection. \textbf{(a)} and \textbf{(d)} Output photon flux (blue curve) and $g^2(0)$ (red curve) versus pump rate. \textbf{(b)} and \textbf{(e)} Quadrature variances $\Delta X^2$ (red dashed curve) and $\Delta Y^2$ (blue solid curve) versus pump rate. \textbf{(c)} and \textbf{(f)} Photon number uncertainty relative to that of coherent light $\Delta n/\sqrt{n}$ (blue curve) and variance product $\Delta X^2\times \Delta Y^2$ (red curve) versus pump rate. The vertical dashed line locates the pump rate giving maximum squeezing and the horizontal dotted lines indicate the quantum limits.}
          \label{fig2}
\end{figure*}

In this section, we present the derivation of a theory for investigating squeezed light generation with semiconductor QDs, such as self-assembled InAs QDs. We consider the experimental configuration depicted in Figure \ref{fig: sketch}a, where QDs are excited by an incoherent pump and subjected to a laser field injected into the cavity. The task is to treat the quantum correlations created by the interaction between the injected field and the excited electron-hole pairs \cite{kreinberg2017emission}. We will compute the electromagnetic field quadratures and look for experimental conditions where nonlinearities produce squeezed light at the output. 

To obtain the equations for the field quadratures, we follow conventional notation and define
\begin{align}
    X &= \frac{1}{2}(a + a^\dagger)
\end{align}

\begin{align}
    Y &= \frac{1}{2i}(a - a^\dagger),
\end{align}

\noindent where $a$ $(a\dagger)$ is the annihilation (creation) operator for a cavity photon. Following this, the variances are: 

\begin{align}
\Delta X^2 &= \langle X^2 \rangle - \langle X \rangle^2 \notag \\
           &= \frac{1}{4}\left[1 + 2\delta\langle a^\dagger a \rangle + 2\text{Re}(\delta\langle aa \rangle)\right], \\
\Delta Y^2 &= \langle Y^2 \rangle - \langle Y \rangle^2 \notag \\
           &= \frac{1}{4}\left[1 + 2\delta\langle a^\dagger a \rangle - 2\text{Re}(\delta\langle aa \rangle)\right],
\end{align}

\begin{figure*}[t] \centering
     \includegraphics[scale=1.2]{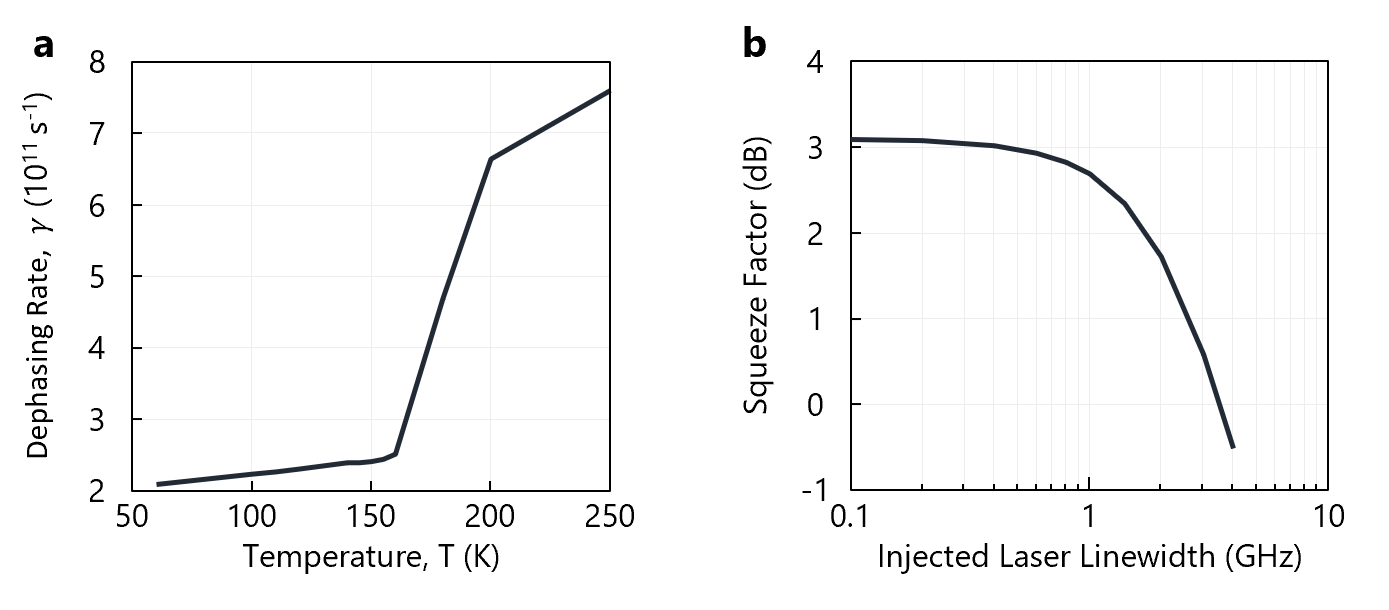}  
          \caption{\textbf{(a)} Dephasing rate ($\gamma (10^{11} s^{-1}$)) versus temperature (T (K)) for InAs QD active medium. \textbf{(b)} Squeeze factor versus linewidth from injected laser frequency drift. All other parameters are similar to those used for Fig. \ref{fig2}.}
          \label{fig3}
\end{figure*}

\noindent where the correlations: $\delta \langle a^\dagger a \rangle = \langle a^\dagger a \rangle - \langle a^\dagger \rangle \langle a \rangle$, and $\delta \langle a a \rangle = \langle a a \rangle - \langle a \rangle \langle a \rangle$.

To completely describe the device performance, we include the photon output rate,
\begin{align}
    P_{out} = 2\gamma_{cav} \langle a^\dagger a \rangle,
\end{align}
 
\noindent where $\gamma_{cav}$ is the cavity decay rate, and the equal temporal second-order correlation function, 

\begin{align}
g^{(2)}(0) &= \frac{\langle a^\dagger a^\dagger a a \rangle}{\langle a^\dagger a \rangle \langle a^\dagger a \rangle}.
\end{align}

\noindent Together, Eqns. (3)-(6) provide complementary insight into the non-classical nature of light. To derive the equations of motion necessary for evaluating Eqns. (3)-(6), the starting point is the following system Hamiltonian:

\begin{align}
    H &= \sum_{\alpha} \varepsilon_{e\alpha} \, c_{\alpha}^\dagger c_{\alpha} 
+ \sum_{\beta} \varepsilon_{h\beta} \, b_{\beta}^\dagger b_{\beta} 
+ \hbar \nu \left(a^\dagger a + \frac{1}{2} \right) \notag \\
&\quad - i \hbar g \sum_{\alpha} 
\left( b_{\alpha}^\dagger c_{\alpha}^\dagger a 
- a^\dagger c_{\alpha} b_{\alpha} \right) 
\label{eq:hamiltonian},
\end{align}

\noindent where $c_\alpha$ $(c_\alpha^\dagger)$ and $b_\alpha$ $(b_\alpha^\dagger)$ are annihilation (creation) operators for electrons and holes respectively, and the summation is over an inhomogeneously broadened QD distribution. The light-matter interaction coupling constant $g$ is then


\begin{align}
    g = \frac{\mathcal{P}}{\hbar} \sqrt{\frac{\hbar \nu}{V \varepsilon_B}},
\end{align}

\noindent where $\mathcal{P}$ is the QD dipole matrix moment, $V$ is the mode volume, and $\varepsilon_B$ is the background permittivity. Using the Hamiltonian, the equation of motion for the expectation value of the photon field operator is

\begin{align}
    \frac{d\langle A \rangle}{dt} = -\gamma_c \langle A \rangle + g \sum_{\alpha} \langle C_{\alpha} B_{\alpha} \rangle +  A_{\text{inj}}  e^{i(\nu - \nu_{\text{inj}})t},
\end{align}

\noindent where $2\gamma_c$ is the cavity linewidth, and we have transformed into a rotating frame where $\langle a \rangle = \langle A \rangle e^{-i\nu t}$, and $\langle c_\alpha b_\alpha \rangle = \langle C_{\alpha}B_{\alpha} \rangle e^{-i\nu t}$. In the last term of Eqn. 9, we include influence from an injected laser field with strength $A_{inj}$ and frequency $v_{inj}$ \cite{haus1984quantum, chow1975line}. Similar derivations give

\begin{align}
\frac{d\langle C_{\alpha} B_{\alpha} \rangle}{dt} 
&= [i(\nu - \omega_{\alpha}) - \gamma] \langle C_{\alpha} B_{\alpha} \rangle \notag \\
&\quad + g \langle A \rangle \left( \langle C_{\alpha}^\dagger C_{\alpha} \rangle + \langle B_{\alpha}^\dagger B_{\alpha} \rangle - 1 \right) \notag \\
&\quad + g \left( \delta\langle C_{\alpha}^\dagger C_{\alpha} A \rangle + \delta\langle B_{\alpha}^\dagger B_{\alpha} A \rangle \right)  \\
\frac{d\langle C_{\alpha}^\dagger C_{\alpha} \rangle}{dt} 
&= -\gamma_{\text{nr}} \langle C_{\alpha}^\dagger C_{\alpha} \rangle 
- \gamma_{\text{nl}} \langle C_{\alpha}^\dagger C_{\alpha} \rangle \langle B_{\alpha}^\dagger B_{\alpha} \rangle \notag \\
&\quad + P \left(1 - \langle C_{\alpha}^\dagger C_{\alpha} \rangle \right) \notag \\
&\quad - 2g\, \text{Re} \left( \langle A \rangle^* \langle C_{\alpha} B_{\alpha} \rangle + \delta\langle A^\dagger C_{\alpha} B_{\alpha} \rangle \right)  \\
\frac{d\langle B_{\alpha}^\dagger B_{\alpha} \rangle}{dt} 
&= -\gamma_{\text{nr}} \langle B_{\alpha}^\dagger B_{\alpha} \rangle 
- \gamma_{\text{nl}} \langle C_{\alpha}^\dagger C_{\alpha} \rangle \langle B_{\alpha}^\dagger B_{\alpha} \rangle \notag \\
&\quad + P \left(1 - \langle B_{\alpha}^\dagger B_{\alpha} \rangle \right) \notag \\
&\quad - 2g\, \text{Re} \left( \langle A \rangle^* \langle C_{\alpha} B_{\alpha} \rangle + \delta\langle A^\dagger C_{\alpha} B_{\alpha} \rangle \right)
\end{align}

\noindent In Eqns. (10)-(12), $\gamma$ is the QD dephasing rate, $\gamma_{nr}$ is the nonradiative carrier decay rate, and $\gamma_{nl}$ is the spontaneous emission rate into non-lasing modes such as other cavity modes and free space. The incoherent pump rate is $P$ and it is multiplied by a factor describing the Pauli blocking condition. The terms denoted with $\delta$ are the correlations beyond the singlet (mean field) level. \newline

\noindent Returning to Eqns. (3) and (4), we note that separation between $\Delta X^2$ and $\Delta Y^2$ (i.e. squeezing), depends on a nonvanishing $\delta \langle A^2 \rangle$. Continuing with the derivation of equations of motion yields: 

\begin{align}
\frac{d\, \delta\langle A^2 \rangle}{dt} 
&= -2\gamma_c\, \delta\langle A^2 \rangle 
+ 2g \sum_{\alpha} \delta\langle C_{\alpha} B_{\alpha} A \rangle \notag \\
&\quad - 2 \langle A \rangle  A_{\text{inj}} 
e^{i(\nu - \nu_{\text{inj}})t} 
\end{align}

\noindent which explicitly indicates contributions from the injected laser field and the correlation $\delta \langle C_\alpha B_\alpha A \rangle$. The latter term, 

\begin{align}
    \delta\langle C_{\alpha} B_{\alpha} A \rangle 
= \langle C_{\alpha} B_{\alpha} A \rangle 
- \langle C_{\alpha} B_{\alpha} \rangle \langle A \rangle ,
\end{align}

\noindent is the quantum-mechanical phase correlation between the electron-hole polarization and the intracavity laser field. Under strong coupling conditions, it is related to the polariton ($\langle C_\alpha B_\alpha \rangle$ is the polarization and $\langle A \rangle$ is the photon). For device practicality, we do not intend to operate in the strong coupling regime. The equations of motion for other contributing correlations are in the Appendix.

\section{Numerical Results}

\begin{figure*}[t] \centering
     \includegraphics[scale=1.2]{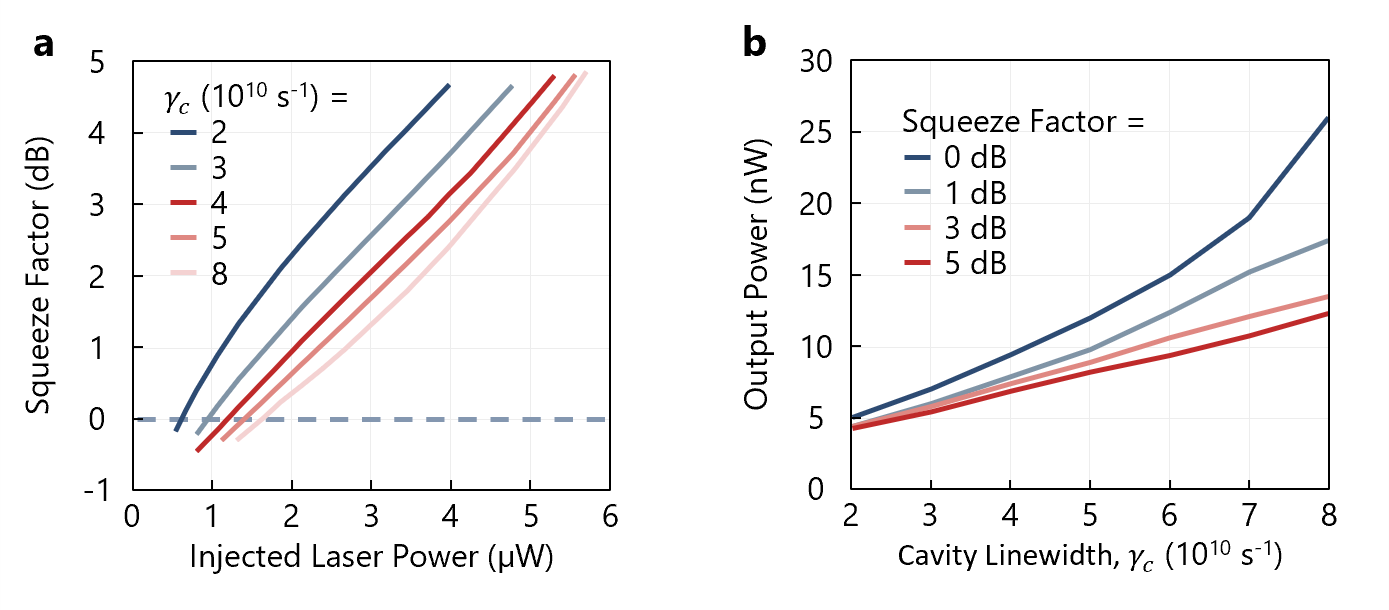}  
          \caption{\textbf{(a)} Squeeze factor versus injected laser power. The curves are for cavity linewidths as indicated. \textbf{(b)} Squeezed light output power versus cavity linewidths ($\gamma_c (10^{10} s^{-1})$). The dark blue curve is for the onset of squeezing. The other curves are for squeeze factors as indicated. All other parameters are similar to those used for Fig. 2. }
          \label{fig4}
\end{figure*}

\noindent Equations~(9)-(12), together with the set (A1)-(A12) in the appendix, are solved numerically for self-assembled InAs QDs positioned at the antinode of a GaAs microcavity mode. We take a QD areal density of $N_{QD} = 2 \times 10^{14}$ $m^{-2}$ and assume an inhomogeneous width of $10$ meV. The microcavity diameter and length are 0.2 $\mu$m and 1.5 $\mu$m, respectively, with linewidth $\gamma_c=2\times10^{10}$ $s^{-1}$. Unless otherwise noted, the calculations use $\lambda =0.92$ $\mu$m, dephasing $\gamma = 2\times 10^{11}$ $s^{-1}$, nonradiative loss $\gamma_{nr} = 2\times 10^{10}$ $s^{-1}$, spontaneous emission into non-lasing modes $\gamma_{nl} = 3 \times 10^{12}$ $s^{-1}$, and dipole matrix element $\mathcal{P}=e \times$0.5 $nm$. These choices are consistent with reported QD and cavity devices \cite{senellart2017high, shang2024ultra}. As sketched in Fig.~\ref{fig: sketch}b, we model an inhomogeneously broadened ensemble of 25 QDs, out of which the calculations indicate 9 are coupled to the cavity optical field.

In Fig. \ref{fig2}, panels (a-c) correspond to a free-running device (no optical injection). Figure~\ref{fig2}a plots the output power (blue) and the zero-delay intensity autocorrelation $g^{(2)}(0)$ (red) versus pump rate, revealing the transition from spontaneous to stimulated emission as the pump increases. The output rises while $g^{(2)}(0)$ evolves from thermal-like values ($\approx 2$) toward the Poisson level ($\approx 1$) characteristic of coherent emission. Figure~\ref{fig2}b reports the quadrature variances $\Delta X^2$ and $\Delta Y^2$. Without injection they coincide at $1/4$, indicating no squeezing and a coherent minimum-uncertainty state. In Fig.~\ref{fig2}c we show the photon-number noise normalized to coherent radiation, $\Delta n/\sqrt{n}$ (blue), together with the variance product $\Delta X^2\times\Delta Y^2$ (red). Above threshold, the curves approach $\Delta n/\sqrt{n}=1$ and $\Delta X^2\times\Delta Y^2=1/16$, as expected for Poisson statistics under our normalization.

Panels (d-f) of Fig. \ref{fig2} include an additional external laser injection at a rate of photon injection, $\mathcal{L}_{inj} = |A_{inj}|^2 /T_{out} = 20$ photons/$ps$, where $T_{out} = n_B L\gamma_c / c$ (see Fig. \ref{fig: sketch}a). The output and $g^{(2)}(0)$ versus pump in Fig. \ref{fig2}d follow trends similar to Fig. \ref{fig2}a, with a modest increase at low pump from the injected photons. In Fig. \ref{fig2}e, the variances separate, $\Delta X^2\neq\Delta Y^2$, indicating the onset of squeezing just before lasing. The curves in Fig. \ref{fig2}f confirm this behavior, at the point of maximum separation, the variance product remains at $1/16$ (minimum-uncertainty), while $\Delta n/\sqrt{n}<1$ indicates sub-Poissonian photon statistics. In practice, amplitude squeezing is maintained only if classical pump relative-intensity noise (RIN) is suppressed below shot noise. The Yamamoto-Machida analysis predicts that a semiconductor laser becomes an intrinsic squeezer once the pump RIN lies beneath the cavity bandwidth \cite{yamamoto1986amplitude}, a conclusion experimentally confirmed under quiet, constant-current operation \cite{zhao2024broadband}. Typical diode sources add on the order of $20$-$30$~dB excess RIN at RF, motivating low-impedance “quiet” drivers, feed-forward cancellation, or active feedback.

A potential concern raised by Fig. \ref{fig2} is sensitivity to the injected-laser frequency stability. Simulations that average over a distribution of injected frequencies (to model finite laser linewidth) show that squeezing remains essentially unaffected so long as the injected linewidth is roughly three orders of magnitude smaller than the dephasing rate. Figure \ref{fig3}a displays the dephasing rate versus temperature for the InAs QD medium under study. Calculations based on quantum-kinetic carrier-carrier and carrier-phonon scattering \cite{chow2013physics} indicate a monotonic decrease with cooling due to the reduced phonon population. For $T<160$ K, electron-phonon scattering is negligible and the dephasing approaches $\gamma\approx 2\times10^{11}$ $s^{-1}$ at the carrier density $N=6\times10^{15}$ $m^{-2}$, corresponding to the squeezing maximum in Fig. \ref{fig2}. Using this $\gamma$, Fig. \ref{fig3}b shows that the squeeze factor, defined as $[-10\log(\Delta n^2/N)]$ \cite{loudon1987squeezed}, is insensitive to slow injected-linewidth broadening up to $\sim$1 GHz, beyond which degradation becomes steep. Although intrinsic DFB/DBR linewidths are typically in the kHz range, slow thermal/mechanical drifts can broaden the effective linewidth over the measurement time and thus limit phase-referenced experiments. This underscores the need for narrow-linewidth sources and good thermo-mechanical stability to preserve quadrature squeezing in the presence of injection-induced phase noise. The model assumes a 10 $meV$ Gaussian inhomogeneous distribution, employing narrower ensembles-for example, pyramidal InGaAs QDs \cite{zhang2022chip}, tightens the collective coupling, reduces the pump current needed to reach the squeezed regime, and broadens the usable squeezing bandwidth.

 Also of interest are the roles of injected power and cavity linewidth. Figure \ref{fig4}a shows that squeezing increases with injected power. The curves for different $\gamma_c$ indicate that the squeeze factor becomes nearly injection-independent when $\gamma_c>5\times10^{10}$ $s^{-1}$. As anticipated, stronger optical injection enhances noise reduction, while smaller cavity decay rates (higher $Q$) enable deeper suppression. For instance, a squeeze factor of about $4$ dB is obtained with $\sim4$ $\mu$W of injected power when $\gamma_c=3\times10^{10}$ $s^{-1}$ ($Q = 6.7 \times 10^{4}$. By comparison, state-of-the-art nonlinear microring resonators typically require $100$-$200$ mW of CW pump to reach the below threshold OPO regime where sub dB squeezing appears \cite{yang2021squeezed}. In contrast, the QD devices analyzed here convert sub mW electrical input into $\mu$W level optical power while achieving comparable quadrature-noise reduction. Finally, Fig. \ref{fig4}b plots squeezed light output power versus cavity linewidth for several target squeeze factors, maintaining stronger squeezing demands narrower effective linewidths (smaller $\gamma_c$), reflecting the familiar trade-off between output power and preservation of quantum correlations.

\section{Conclusions}

We developed a cavity-QED framework for an inhomogeneously broadened QD ensemble that is incoherently pumped and weakly driven by a coherent seed, solved for steady-state operation, and evaluated the output-field quadrature statistics. The model identifies practical operating points where the amplitude-quadrature variance dips below the Poissonian (coherent-state) bound, with squeezing levels approaching $\sim$5 dB under device parameters already demonstrated in QD-microcavity platforms. We also find that four-wave-mixing induced quantum correlations simultaneously set the squeezed/anti-squeezed noise ellipse and reshape the small-signal gain spectrum, providing a quantum counterpart to carrier-mediated mean-field interactions known to induce self-mode locking and the very narrow linewidths observed under self-injection locking.

From an engineering perspective, three levers govern performance, cavity decay, seed strength, and dephasing. Lower $\gamma_c$ deepens noise suppression, stronger optical injection increases the squeezing until cavity loss dominates, and reduced dephasing (via temperature and material choice) broadens the regime where squeezing is robust. Consistent with the Yamamoto-Machida picture, amplitude squeezing persists only when pump relative-intensity noise is pushed below the cavity bandwidth, a condition met by quiet, constant-current drive and verified in recent experiments. Compared with $\chi^{(2)}/\chi^{(3)}$ benchtop systems, the QD platform achieves comparable noise reduction with $\mu$W level optical powers and chip-scale footprints.

Looking ahead, the same formalism extends naturally to optimized-quadrature detection, pulsed operation, and multimode cavities, as well as to ensembles with narrower inhomogeneous widths to tighten collective coupling and lower pump requirements. Integrating phase-stabilized local oscillators, low-RIN current drivers, and feedback for slow frequency drift should enable stable, room-temperature squeezed light sources co-fabricated with classical control electronics and waveguide networks, supporting quantum communications, sensing, and information processing at scale. This model can be further improved upon by including the numerical expansion for triplet correlations.

\begin{acknowledgments}
\noindent The authors acknowledge the Centre for Integrated Nanotechnologies, an Office of Science User Facility operated for the U.S. Department of Energy (DOE) Office of Science by Sandia National Laboratories. The research at UCSB was supported by the NSF Quantum Foundry through the Q-AMASE-i Program (Grant No. DMR-1906325), the NSF CAREER Program (Grant No. 2045246), and the Defense Advanced Research Projects Agency (Award No. D24AC00166-00). Frederic Grillot also acknowledges the discovery funding from the Natural Sciences and Engineering Research Council of Canada (NSERC).

\end{acknowledgments}

\allowdisplaybreaks[1]
\section*{Appendix}
\noindent This appendix lists the equations of motion for the correlations appearing on the right-hand side of Eqns. (10)-(13).

\begin{align}
&\frac{d\,\delta\langle C_{\alpha}B_{\alpha}A\rangle}{dt}
 = \bigl[i(\nu-\omega_{\alpha})-(\gamma_c+\gamma)\bigr]
   \delta\langle C_{\alpha}B_{\alpha}A\rangle
\notag \\
&-g\langle C_{\alpha} B_{\alpha} \rangle^2 - g\,\langle C_{\alpha}B_{\alpha}\rangle A_{\text{inj}}
   e^{i(\nu-\nu_{\text{inj}})t} 
\notag \\
&+ 2g\,\langle A\rangle
   \bigl[\delta\langle C_{\alpha}^{\dagger}C_{\alpha}A\rangle
        + \delta\langle B_{\alpha}^{\dagger}B_{\alpha}A\rangle\bigr]
\notag \\
&+ g\,\delta\langle A^{2}\rangle
   \bigl[\langle C_{\alpha}^{\dagger}C_{\alpha}\rangle
        + \langle B_{\alpha}^{\dagger}B_{\alpha}\rangle - 1\bigr]
\notag \\
&+ g\sum_{\beta} \delta\langle C_{\alpha}B_{\alpha}C_{\beta}B_{\beta}\rangle
\tag{A1}
\end{align}

\begin{align}
&\frac{d\,\delta\langle C_{\alpha}^{\dagger}C_{\alpha}A\rangle}{dt}
= -(\gamma_c+\gamma_{\text{nr}})
   \delta\langle C_{\alpha}^{\dagger}C_{\alpha}A\rangle
\notag\\
&- g\,\langle C_{\alpha}^{\dagger}C_{\alpha}\rangle
     \langle C_{\alpha}B_{\alpha}\rangle
\notag\\
&- g\,\langle C_{\alpha}^{\dagger}C_{\alpha}\rangle
      A_{\text{inj}}
     e^{i(\nu-\nu_{\text{inj}})t}
\notag\\
&- g\!\bigl[\langle C_{\alpha}B_{\alpha}\rangle^{*}\delta\langle A^{2}\rangle
     + \langle C_{\alpha}B_{\alpha}\rangle\,\delta\langle A^{\dagger}A\rangle\bigr]
\notag\\
&- g\!\bigl[\langle A\rangle\,\delta\langle A^{\dagger}C_{\alpha}B_{\alpha}\rangle^{*}
     + \langle A\rangle^{*}\delta\langle C_{\alpha}B_{\alpha}A\rangle\bigr]
\notag\\
&+ g\sum_{\beta\neq\alpha} \delta\langle C_{\alpha}^{\dagger}C_{\beta}B_{\beta}C_{\alpha}\rangle
\tag{A2}
\end{align}

\begin{align}
&\frac{d\,\delta\langle B_{\alpha}^{\dagger}B_{\alpha}A\rangle}{dt}
 = -(\gamma_c+\gamma_{\text{nr}})
   \delta\langle B_{\alpha}^{\dagger}B_{\alpha}A\rangle
\notag\\
&- g\,\langle B_{\alpha}^{\dagger}B_{\alpha}\rangle
      \langle C_{\alpha}B_{\alpha}\rangle
\notag\\
&- g\,\langle B_{\alpha}^{\dagger}B_{\alpha}\rangle
       A_{\text{inj}}
      e^{i(\nu-\nu_{\text{inj}})t}
\notag\\
&- g\!\bigl[
      \langle C_{\alpha}B_{\alpha}\rangle^{*}\delta\langle A^{2}\rangle
      + \langle C_{\alpha}B_{\alpha}\rangle\,\delta\langle A^{\dagger}A\rangle
      \bigr]
\notag\\
&- g\!\bigl[
      \langle A\rangle\,\delta\langle A^{\dagger}C_{\alpha}B_{\alpha}\rangle ^{*}
      + \langle A\rangle^{*}\delta\langle C_{\alpha}B_{\alpha}A\rangle
      \bigr]
\notag\\
&+ g\sum_{\beta\neq\alpha}\delta\langle B_{\alpha}^{\dagger}C_{\beta}B_{\beta}B_{\alpha}\rangle
\tag{A3}
\end{align}

\begin{align}
&\frac{d\,\delta\langle A^{\dagger}C_{\alpha}B_{\alpha}\rangle}{dt}
 = \bigl[i(\nu-\omega_{\alpha})-(\gamma_c+\gamma)\bigr]\,
   \delta\langle A^{\dagger}C_{\alpha}B_{\alpha}\rangle
\notag\\
&+ g\,\langle C_{\alpha}^{\dagger}C_{\alpha}\rangle
      \langle B_{\alpha}^{\dagger}B_{\alpha}\rangle
\notag\\
&+ g\,\langle A\rangle
      \bigl[\delta\langle C_{\alpha}^{\dagger}C_{\alpha}A\rangle^{*}
            + \delta\langle B_{\alpha}^{\dagger}B_{\alpha}A\rangle^{*}\bigr]
\notag\\
&+ g\,\delta\langle A^{\dagger}A\rangle
      \bigl[\langle C_{\alpha}^{\dagger}C_{\alpha}\rangle
            + \langle B_{\alpha}^{\dagger}B_{\alpha}\rangle - 1\bigr]
\notag\\
&+ g\sum_{\beta}\delta\langle B_{\beta}^{\dagger}C_{\beta}^{\dagger}
                    C_{\alpha}B_{\alpha}\rangle
\notag\\
&- g\,\langle C_{\alpha}B_{\alpha}\rangle
      A_{\text{inj}}^{*}
      e^{-i(\nu-\nu_{\text{inj}})t}
\tag{A4}
\end{align}

\begin{align}
&\frac{d\,\delta\langle C_{\alpha}B_{\alpha}C_{\beta}B_{\beta}\rangle}{dt}
 = \bigl[-2\gamma + i\,(2\nu-\omega_{\alpha}-\omega_{\beta})\bigr]\,
   \delta\langle C_{\alpha}B_{\alpha}C_{\beta}B_{\beta}\rangle
\notag\\
&+ g\bigl[\langle C_{\alpha}^{\dagger}C_{\alpha}\rangle
        + \langle B_{\alpha}^{\dagger}B_{\alpha}\rangle - 1\bigr]\,
   \delta\langle C_{\beta}B_{\beta}A\rangle
\notag\\
&+ g\bigl[\langle C_{\beta}^{\dagger}C_{\beta}\rangle
        + \langle B_{\beta}^{\dagger}B_{\beta}\rangle - 1\bigr]\,
   \delta\langle C_{\alpha}B_{\alpha}A\rangle
\notag\\
&+ g\,\langle A\rangle\,
   \Bigl[\delta\langle C_{\alpha}^{\dagger}C_{\beta}B_{\beta}C_{\alpha}\rangle
        + \delta\langle B_{\alpha}^{\dagger}C_{\beta}B_{\beta}B_{\alpha}\rangle\Bigr]
        \notag\\
&+ g\,\langle A\rangle\,
   \Bigl[\delta\langle C_{\beta}^{\dagger}C_{\alpha}B_{\alpha}C_{\beta}\rangle
        + \delta\langle B_{\beta}^{\dagger}C_{\alpha}B_{\alpha}B_{\beta}\rangle\Bigr]
\tag{A5}
\end{align}

\vspace{4pt}
\begin{align}
&\frac{d\,\delta\langle C_{\alpha}^{\dagger}C_{\beta}B_{\beta}C_{\alpha}\rangle}{dt}
 = \bigl[-(\gamma_{\text{nr}}+\gamma)+i(\nu-\omega_{\beta})\bigr]\,
   \delta\langle C_{\alpha}^{\dagger}C_{\beta}B_{\beta}C_{\alpha}\rangle
\notag\\
&+ g\bigl[\langle C_{\beta}^{\dagger}C_{\beta}\rangle
         + \langle B_{\beta}^{\dagger}B_{\beta}\rangle - 1\bigr]\,
   \delta\langle C_{\alpha}^{\dagger}C_{\alpha}A\rangle
\notag\\
&- g\bigl[ \langle C_{\alpha}B_{\alpha}\rangle^{*}
       \delta\langle C_{\beta}B_{\beta}A\rangle
+ \langle C_{\alpha}B_{\alpha}\rangle
       \delta\langle A^{\dagger}C_{\beta}B_{\beta}\rangle \bigr]
\notag\\
&+ g\,\langle A\rangle \bigl[
       \delta\langle C_{\beta}^{\dagger}B_{\beta}^{\dagger}B_{\beta}C_{\alpha}\rangle
+ 
       \delta\langle C_{\alpha}^{\dagger}C_{\beta}^{\dagger}C_{\beta}C_{\alpha}\rangle
\notag\\
&- 
       \delta\langle B_{\beta}^{\dagger}C_{\beta}^{\dagger}C_{\alpha}B_{\alpha}\rangle ^{*} \bigr]
- \langle A\rangle^{*}\,
     \delta\langle C_{\alpha}B_{\alpha}C_{\beta}B_{\beta}\rangle
\tag{A6}
\end{align}

\vspace{4pt}
\begin{align}
&\frac{d\,\delta\langle B_{\alpha}^{\dagger}C_{\beta}B_{\beta}B_{\alpha}\rangle}{dt}
 = \bigl[-(\gamma_{\text{nr}}+\gamma)+i(\nu-\omega_{\beta})\bigr]\,
   \delta\langle B_{\alpha}^{\dagger}C_{\beta}B_{\beta}B_{\alpha}\rangle
\notag\\
&+ g\bigl[\langle C_{\beta}^{\dagger}C_{\beta}\rangle
        + \langle B_{\beta}^{\dagger}B_{\beta}\rangle - 1\bigr]\,
   \delta\langle B_{\alpha}^{\dagger}B_{\alpha}A\rangle
\notag\\
&- g\
   \Bigl[\langle C_{\alpha}B_{\alpha}\rangle^{*}
         \delta\langle C_{\beta}B_{\beta}A\rangle
        + \langle C_{\alpha}B_{\alpha}\rangle \delta \langle A^{\dagger}C_{\beta}B_{\beta} \rangle \Bigr]
\notag\\
&+ g \langle A \rangle \bigl[ \delta \langle C_{\beta}^{\dagger}B_{\alpha}^{\dagger}B_{\alpha}C_{\beta} \rangle + \delta \langle B_{\alpha}^{\dagger}B_{\beta}^{\dagger}B_{\beta}B_{\alpha} \rangle - \delta \langle B_{\alpha}^{\dagger} C_{\alpha}^{\dagger}C_{\beta}B_{\beta} \rangle \bigr]
\notag\\
&- g\,\langle A\rangle^{*}\,
   \delta\langle C_{\alpha}B_{\alpha}C_{\beta}B_{\beta}\rangle
\tag{A7}
\end{align}

\vspace{4pt}
\begin{align}
&\frac{d\,\delta\langle C_{\alpha}^{\dagger}B_{\alpha}^{\dagger}
   B_{\alpha}C_{\alpha}\rangle}{dt}
 = -2\gamma_{\text{nr}}\,
   \delta\langle C_{\alpha}^{\dagger}
   B_{\alpha}^{\dagger}B_{\alpha}C_{\alpha}\rangle
\notag\\
&+ 2g\bigl[\langle C_{\alpha}^{\dagger}C_{\alpha}\rangle
         + \langle B_{\alpha}^{\dagger}B_{\alpha}\rangle - 1\bigr]\,
   \operatorname{Re}\!\bigl[
        \delta\langle A^{\dagger}C_{\alpha}B_{\alpha}\rangle\bigr]
\notag\\
&- 2g\,\operatorname{Re}\!\Bigl[
        \langle C_{\alpha}B_{\alpha}\rangle^{*}\!
        \bigl(\delta\langle C_{\alpha}^{\dagger}C_{\alpha}A\rangle
             + \delta\langle B_{\alpha}^{\dagger}B_{\alpha}A\rangle\bigr)\Bigr]
\tag{A8}
\end{align}

\vspace{4pt}
\begin{align}
&\frac{d\,\delta\langle
   C_{\alpha}^{\dagger}B_{\beta}^{\dagger}B_{\beta}C_{\alpha}\rangle}{dt}
 = -2\gamma_{\text{nr}}\,
   \delta\langle C_{\alpha}^{\dagger}B_{\beta}^{\dagger}
   B_{\beta}C_{\alpha}\rangle
\notag\\
&- 2g\,\operatorname{Re}\!\Bigl[
     \langle C_{\alpha}B_{\alpha}\rangle^{*}\,
       \delta\langle B_{\beta}^{\dagger}B_{\beta}A\rangle
     + \langle C_{\beta}B_{\beta}\rangle^{*}\,
       \delta\langle C_{\alpha}^{\dagger}C_{\alpha}A\rangle\Bigr]
\notag\\
&- 2g\,\operatorname{Re}\!\Bigl[
     \langle A\rangle^{*}\,
     \bigl(\delta\langle C_{\alpha}^{\dagger}C_{\beta}B_{\beta}C_{\alpha}\rangle
          + \delta\langle B_{\beta}^{\dagger}C_{\alpha}B_{\alpha}C_{\beta}\rangle\bigr)
     \Bigr]
\tag{A9}
\end{align}

\vspace{4pt}
\begin{align}
&\frac{d\,\delta\langle
   C_{\alpha}^{\dagger}C_{\beta}^{\dagger}C_{\beta}C_{\alpha}\rangle}{dt}
 = -2\gamma_{\text{nr}}\,
   \delta\langle C_{\alpha}^{\dagger}C_{\beta}^{\dagger}
               C_{\beta}C_{\alpha}\rangle
\notag\\
&- 2g\,\operatorname{Re}\!\Bigl[
     \langle C_{\alpha}B_{\alpha}\rangle^{*}\,
       \delta\langle C_{\beta}^{\dagger}C_{\beta}A\rangle
     + \langle C_{\beta}B_{\beta}\rangle^{*}\,
       \delta\langle C_{\alpha}^{\dagger}C_{\alpha}A\rangle\Bigr]
\notag\\
&- 2g\,\operatorname{Re}\!\Bigl[
     \langle A\rangle^{*}\,
     \bigl(\delta\langle C_{\alpha}^{\dagger}C_{\beta}B_{\beta}C_{\alpha}\rangle
          + \delta\langle C_{\beta}^{\dagger}C_{\alpha}B_{\alpha}C_{\beta}\rangle\bigr)
     \Bigr]
\tag{A10}
\end{align}


\begin{align}
&\frac{d\,\delta\langle
  B_{\alpha}^{\dagger}B_{\beta}^{\dagger}B_{\beta}B_{\alpha}\rangle}{dt}
 = -2\gamma_{\text{nr}}\,
   \delta\langle B_{\alpha}^{\dagger}B_{\beta}^{\dagger}
   B_{\beta}B_{\alpha}\rangle
\notag\\
&- 2g\,\operatorname{Re}\!\Bigl[
     \langle C_{\alpha}B_{\alpha}\rangle^{*}\,
       \delta\langle B_{\beta}^{\dagger}B_{\beta}A\rangle
     + \langle C_{\beta}B_{\beta}\rangle^{*}\,
       \delta\langle B_{\alpha}^{\dagger}B_{\alpha}A\rangle
     \Bigr]
\notag\\
&- 2g\,\operatorname{Re}\Bigl[ \langle A \rangle^{*}(\delta \langle B_{\beta}^{\dagger}C_{\alpha}B_{\alpha}B_{\beta}\rangle + \delta \langle B_{\alpha}^{\dagger}C_{\beta}B_{\beta}B_{\alpha} \rangle ) \Bigr]
\tag{A11}
\end{align}


\begin{align}
&\frac{d\,\delta\langle A^{\dagger}A\rangle}{dt}
 = -2\gamma_c\,\delta\langle A^{\dagger}A\rangle
+ 2g\sum_{\alpha}
   \operatorname{Re}\!\bigl[\delta\langle A^{\dagger}C_{\alpha}B_{\alpha}\rangle\bigr]
\notag\\
&- 2\,\operatorname{Re}\!\Bigl[
   \langle A\rangle A_{\text{inj}}^{*}
   e^{-i(\nu-\nu_{\text{inj}})t}\Bigr]
\tag{A12}
\end{align}

For further understanding, it is instructive to formally integrate Eqns. (10)-(12) and (A1)-(A3), followed by a perturbation treatment around the saturated carrier density. The process gives a 2$^{\text{nd}}$ order correction to the carrier populations that is proportional to intensity $\langle A^\dagger \rangle \langle A \rangle$. More algebra shows the correlations $\delta \langle C_{\alpha}^\dagger C_\alpha A \rangle$ and $\delta \langle B_{\alpha}^\dagger B_\alpha A \rangle$, which are important for squeezing, have the leading terms proportional to $\langle A^\dagger \rangle \langle A \rangle \langle A \rangle$, thus implying a four-wave mixing origin.
\newpage

\bibliography{Biblio}

\end{document}